\documentclass[12pt]{article}
\usepackage{latexsym,amsfonts,amssymb}
\title{Canonically conjugate variables for the 
periodic Camassa-Holm equation}
\author{Alexei V. Penskoi\thanks{Centre de Recherches Math\'ematiques, 
Universit\'e de Montr\'eal,
C.~P.~6128, Succ. Centre-ville, Montr\'eal, 
Qu\'ebec, H3C 3J7, Canada {\tt e-mail: penskoi@crm.umontreal.ca}}
\thanks{Current address: Independent University of Moscow,
Bolshoy Vlasyevskiy Pereulok 11, 119002 Moscow, Russia
{\tt e-mail: penskoi@mccme.ru}}}
\date{}
\begin{document}
\maketitle

\abstract{The Camassa-Holm shallow water equation is known to be Hamiltonian 
with respect to two compatible Poisson brackets.
A set of conjugate variables is constructed for both brackets
using spectral theory.}

\smallskip


\smallskip


\section{Introduction}
In 1976 Flaschka and McLaughlin observed~\cite{FM} on particular examples of
the Korteweg-de~Vries equation and the Toda lattice with periodic
boundary conditions that variables arising naturally from spectral theory 
and algebraic geometry have ``nice'' symplectic properties. It was surprising
because a priori it is not clear why there 
should exist a relation between the Lax pairs and 
the Hamiltonian formalism of the corresponding equations.
The same phenomenon was later observed on numerous examples. 
This led Novikov and Veselov to the theory of algebro-geometric Poisson 
brackets on the universal bundle of hyperelliptic curves~\cite{VN1,VN2}.
Later Krichever and Phong developed~\cite{KP1,KP2} a unified construction
of the symplectic forms arising in the $N=2$ Yang-Mills theories and
soliton equations. 

The goal of this paper is to prove analogues of some results of Flaschka and 
McLaughlin~\cite{FM} for the Camassa-Holm equation, also known as the shallow
water equation:
\begin{equation}\label{CH1}
v_t-v_{xxt}+3vv_x-2v_xv_{xx}-vv_{xxx}=0.
\end{equation}

The Camassa-Holm equation is known to be bi-Hamiltonian~\cite{CH}.
To describe two compatible Poisson structures, it is better to use
the function $m=(1-D^2)v$ instead of $v.$ Here and later 
$D$ and ${}'$ denote the derivative with respect to $x.$
Let $J=mD+Dm$ and $K=\frac{1}{2}D(1-D^2).$
Then (in the periodic case, i.e. $v(x+1)=v(x)$) 
the two compatible Poisson brackets are given by the formulae
$$
\{A,B\}_1=\int_0^1\frac{\partial A}{\partial m}J%
\frac{\partial B}{\partial m}\,dx
$$
and
$$
\{A,B\}_2=\int_0^1\frac{\partial A}{\partial m}K%
\frac{\partial B}{\partial m}\,dx.
$$
The Camassa-Holm equation is Hamiltonian with respect to both brackets,
it can be rewritten as
$$
m_t+\{m,H_2\}_1=0\quad\mbox{or}\quad m_t+\{m,H_3\}_2=0,
$$
where $H_2=\frac{1}{2}\int_0^1(v^2+(v')^2)\,dx$ and
$H_3=\int_0^1(v^3+v(v')^2)\,dx.$

The Camassa-Holm equation can be expressed as a compatibility 
condition of two equations ~\cite{CH}. 
Following~\cite{CM}, we will write these equations as
\begin{equation}\label{Sp}
\psi''=\frac{1}{4}\psi -\lambda m\psi,
\end{equation}
$$
\psi_t=-\left(v+\frac{1}{2\lambda}\right)\psi'+\frac{1}{2}v'\psi.
$$

The Camassa-Holm equation has some particular properties. 
Firstly, it is necessary to consider not only
smooth $m$ but also distributions. Indeed, even for the traveling
wave solution $u(x,t)=ce^{-|x-ct|}$ we have $m=2c\delta(x-ct).$ It creates
some difficulties since solutions $\psi$ of
equation~(\ref{Sp}) are not smooth.
The corresponding solutions of~(\ref{CH1}) are not classical but
weak solutions, see~\cite{CE,CMo} for a discussion.
Secondly, the dynamics is not linear on the Jacobian of the spectral curve, 
it is necessary to consider some covering of the spectral curve. It implies
that one should use an analogue of the Abel map using meromorphic differentials
to linearize the dynamics. As a result, one has to use piecewise
meromorphic functions to write down
algebro-geometric solutions. One can find more
details in~\cite{CM, ACFHM}. The algebro-geometric solutions are also studied
in~\cite{GH}.

Since the application of the available general theories from
the papers~\cite{VN1,VN2,KP1,KP2} is not obvious because of 
these particularities, we use the methods
similar to those of Flaschka and McLaughlin~\cite{FM}. 

Let us now recall some results of~\cite{FM}. The Korteweg-de~Vries equation
$$
u_t-6uu_x+u_{xxx}=0
$$
is known to be Hamiltonian with respect to the Poisson bracket given
(in the periodic case, i.e.  $u(x+1)=u(x)$) by the formula 
$$
\{A,B\}_{\mbox{\tiny KdV}}=\int_0^1\frac{\partial A}{\partial m}D%
\frac{\partial B}{\partial m}\,dx.
$$
Let us consider the spectral problem for the Schr\"odinger operator
\begin{equation}\label{Sch}
-y''+uy=\lambda y.
\end{equation}
Let $y_2(x,\lambda)$ be a solution of~(\ref{Sch}) normalized by the conditions
$y_2(0,\lambda)=0,$ $y_2'(0,\lambda)=1.$
Auxiliary eigenvalues $\mu_i$ are solutions of the equation $y_2(1,\mu)=0.$
The functions $y_2(x,\mu_i)$ are Floquet solutions of the spectral 
problem~(\ref{Sch}), i.e. there exists Floquet multipliers
$\rho_i$ such that $y_2(x+1,\mu_i)=\rho_i y_2(x,\mu_i).$

Flaschka and McLaughlin proved~\cite{FM} that $\mu_i$ and
$f_j=-2\log|\rho_j|$ are conjugate variables, i.e.
$$
\{\mu_i,\mu_j\}_{\mbox{\tiny KdV}}=0,\quad
\{\mu_i,f_j\}_{\mbox{\tiny KdV}}=\delta_{ij},\quad
\{f_i,f_j\}_{\mbox{\tiny KdV}}=0.
$$

In this paper we prove analogues of this result for the
Camassa-Holm equation. Let $\mu_i$ and $\rho_j$ be now auxiliary eigenvalues
and corresponding Floquet multipliers for the spectral problem~(\ref{Sp}).
Using the theory of the spectral problem~(\ref{Sp})
developed by Constantin and McKean~\cite{CM} we prove the
following theorems. 

\noindent{\bf Theorem 1.} The variables $\mu_i$
and $f_j=-\frac{\log|\rho_j|}{\mu_j^2}$ are conjugate with respect to the
first bracket $\{\,,\}_1.$ 

\noindent{\bf Theorem 2.} The variables $\mu_i$ and 
$g_j=-\frac{\log|\rho_j|}{\mu_j^3}$ 
are conjugate with respect to the second bracket $\{\,,\}_2.$

The plan of the paper is as following. In Section~\ref{st} we recall
necessary for us results of Constantin and McKean~\cite{CM} concerning
the theory of the spectral problem~(\ref{Sp}). Then in Section~\ref{proof}
we prove Theorems~1 and 2.

\section{Spectral theory related to the Camassa-Holm equation}\label{st}

\noindent{\bf Definition} We say that a function 
$f:\mathbb{R}\longrightarrow\mathbb{R}$ is piecewise-smooth if
the following conditions hold:
\begin{enumerate}
\item The function $f$ is continuous.
\item For any finite interval $[b,c]$ there exists a
finite number of points $b\leqslant a_1<\dots<a_n\leqslant c$ such that
\begin{enumerate}
\item $f$ is a smooth function on $[b,c]$ except points $a_i$;
\item left and right limits $f'_-(a_i)$ and
$f'_+(a_i)$ of the derivative at the points $a_i$ exist.
\end{enumerate}
\end{enumerate}

In the rest of this paper we assume that $m$ 
can be written as a smooth
function $m_s$ plus a linear combination of delta-functions
$$
m(x,t)=m_s(x,t)+\sum_{n}p_n(t)\delta(x-q_n(t))
$$
such that for any $t$ the set  $\{q_n(t)\}$ has no accumulation points.

This assumption on $m$ is reasonable since on the one hand this includes
both the case of a smooth $m$ and the case of multipeakons
$\sum_{n}p_n(t)\delta(x-q_n(t))$ which are of particular interest. On 
another hand, for such $m$ the Cauchy problem for ODE~(\ref{Sp})
always has a unique solution in the class of piecewise-smooth functions
defined above. Moreover, many usual properties hold, for example the
Wronskian of two solutions is a constant.

Let us recall some results about the spectral problem~(\ref{Sp}) which will be
useful for us. Our basic source is paper of Constantin and McKean~\cite{CM}.
We consider the periodic case, so $v(x)=v(x+1)$ and, respectively, 
$m(x+1)=m(x).$

Let $y_1(x,\lambda)$ and $y_2(x,\lambda)$ be a fundamental set
of solutions of~(\ref{Sp}) defined by normalization
$$
y_1(0,\lambda)=1,\quad y'_1(0,\lambda)=0,
$$
$$
y_2(0,\lambda)=0,\quad y'_2(0,\lambda)=1.
$$

Any solution $\psi$ of~(\ref{Sp}) can be written as linear combination of
$y_1$ and $y_2:$ 
\begin{equation}\label{razl}
\psi(x,\lambda)=\psi(0,\lambda)y_1(x,\lambda)+\psi'(0,\lambda)y_2(\lambda).
\end{equation}
It follows that we have the formula
\begin{equation}\label{transfer}
\left(%
\begin{array}{c}
\psi(x,\lambda)\\
\psi'(x,\lambda)
\end{array}%
\right)=\left(%
\begin{array}{cc}
y_1(x,\lambda)&y_2(x,\lambda)\\
y'_1(x,\lambda)&y'_2(x,\lambda)
\end{array}%
\right)\left(%
\begin{array}{c}
\psi(0,\lambda)\\
\psi'(0,\lambda)
\end{array}%
\right).
\end{equation}
We will denote the matrix from~(\ref{transfer}) by $U(x,\lambda).$

A solution $\psi$ of~(\ref{Sp}) is said to be a Floquet solution
if there exist a number $\rho$ called a Floquet multiplier such that
$$
\psi(x+1,\lambda)=\rho\psi(x,\lambda).
$$
It follows from~(\ref{transfer}) that a Floquet solution is an
eigenvector of $U(1,\lambda)$ and $\rho$ is an eigenvalue of $U(1,\lambda).$
The determinant of $U(x,\lambda)$ is a Wronskian and it is easy to see
from the definition of $y_1$ and $y_2$ that it is equal to $1.$ Hence,
we obtain the following equation for $\rho:$
\begin{equation}\label{rhoeq}
\rho^2+2\Delta(\lambda)\rho+1=0,
\end{equation}
where 
$$
\Delta(\lambda)=\frac{1}{2}\mbox{tr}\,U(1,\lambda)=%
\frac{1}{2}(y_1(1,\lambda)+y'_2(1,\lambda)).
$$

The case $\rho=\pm 1$ corresponds to periodic/antiperiodic solutions.
The corresponding eigenvalues $\lambda^\pm_i$ define a spectral curve.

Let us consider auxiliary eigenvalues $\mu_i$ defined as solutions of
the equation $y_2(1,\mu)=0.$ Since $m(x)$ is periodic, $y_2(x+1,\mu_i)$ is a 
solution of~(\ref{Sp}) for $\lambda=\mu_i.$ Using~(\ref{razl}) we see that
it is proportional to $y_2(x,\mu_i)$ and the proportionality constant is 
$y_2'(1,\mu_i).$ Thus $y_2(x+1,\mu_i)=y'_2(1,\mu_i)y_2(x,\mu_i)$ and this means
that $y_2(x,\mu_i)$ is a Floquet solution with the Floquet multiplier
$\rho_i=y'_2(1,\mu_i).$

Now let us consider the equation~(\ref{rhoeq}) for $\lambda=\mu_i.$ We found
one root of this equation $\rho_i.$ But there is another root 
$\tilde{\rho}_i=1/\rho_i.$ Let $y(x,\mu_i)$ be the corresponding Floquet
solution normalized by the condition $y(0,\mu_i)=1,$ this
normalization is possible since
$y$ and $y_2$ are linearly independent.

The disposition of spectra is very similar to the KdV case: the 
periodic/antiperiodic eigenvalues $\lambda^\pm_i$ define gaps containing
each only one auxiliary eigenvalue $\mu_i.$

Flaschka and McLaughlin~\cite{FM} used in their proofs
identities with Wronskians, but these identities 
are not useful in the case of the Camassa-Holm equation.
It is the identity from the following lemma that will be our main tool.
It was used in~\cite{CM} to calculate Poisson brackets.

\noindent{\bf Lemma.} Let $\psi$ and $\varphi$ be solutions 
(not necessarily different) of the spectral 
problem~(\ref{Sp}) for the same $\lambda.$ Then we have the following
identity:
$$
\lambda J\psi\varphi=K\varphi\psi,
$$
where $J=mD+Dm$ and $K=\frac{1}{2}D(1-D^2)$ are the operators 
used to define Poisson brackets above.

\noindent{\bf Proof} is a direct calculation, one can
write down $K\varphi\psi$ and then eliminate all second and third derivatives
of $\psi$ and $\varphi$ using~(\ref{Sp}) and the derivative of~(\ref{Sp})
with respect to $x.$

\section{Spectral theory and conjugate variables}\label{proof}

Let us consider the auxiliary eigenvalues $\mu_i.$ It is easy to see from
the spectral problem~(\ref{Sp}) that $\mu_i\ne0.$ Thus we can define
variables $f_j=-\frac{\log|\rho_j|}{\mu_j^2}$ and 
$g_j=-\frac{\log|\rho_j|}{\mu_j^3}.$ It should be remarked that we use
$|\rho_i|$ instead of $\rho_i$ only to obtain real-valued $f_j$ and
$g_j.$ If we consider the complex case, we can drop the absolute value
signs, the commutation relations will be the same.

\noindent{\bf Theorem 1.} The variables $\mu_i$ and $f_j$ are conjugate 
with respect to the first bracket:
$$
\{\mu_i,\mu_j\}_1=0,\quad
\{\mu_i,f_j\}_1=\delta_{ij},\quad
\{f_i,f_j\}_1=0.
$$

\noindent{\bf Proof.} Let us start by calculating 
$\frac{\partial\mu_i}{\partial m}.$ We have
\begin{equation}\label{y2}
y_2''(x,\mu_i)=\frac{1}{4}y_2(x,\mu_i)-\mu_imy_2(x,\mu_i).
\end{equation}
Let us now write for simplicity $y_2$ instead of $y_2(x,\mu_i)$
and $\mu$ instead of $\mu_i.$
The variation of~(\ref{y2}) equals to
\begin{equation}\label{dy2}
\delta y_2''=\frac{1}{4}\delta y_2-\delta\mu my_2-%
\mu\delta my_2-\mu m\delta y_2.
\end{equation}
Note that under variation $\mu$ remains an auxiliary eigenvalue.
Let us multiply the
previous identity by $y_2$ and integrate. On the L.H.S.
we should integrate twice by parts and use the identities
$$
y_2(0,\mu)=y_2(1,\mu)=0,\quad\delta y_2(0,\mu)=\delta y_2(1,\mu)=0
$$
and~(\ref{y2}). We obtain on the L.H.S.
$$
\int_0^1\delta y_2(\frac{1}{4}y_2-\mu my_2)\,dx.
$$
Cancelling the same integrals on the L.H.S. and on the R.H.S. we obtain
\begin{equation}\label{dm}
0=-\delta\mu\int_0^1my_2^2\,dx-\int_0^1\delta m\mu y_2^2.
\end{equation}
Let us now remark that $\int_0^1my_2^2\,dx\ne0.$ Indeed, let us 
multiply~(\ref{y2}) by $y_2$ and integrate, we obtain
$$
\int_0^1y_2''y_2=\frac{1}{4}\int_0^1y_2^2-\mu\int_0^1my_2^2.
$$
Using integration by parts we see that
$$
\mu\int_0^1my_2^2=\int_0^1\left[\left(\frac{y_2}{2}\right)^2+(y_2')^2\right]%
\,dx\ne0.
$$
As we remarked before, auxiliary eigenvalue $\mu\ne0,$ hence
$\int_0^1my_2^2\,dx\ne0.$
Thus, we obtain from~(\ref{dm}) that
$$
\frac{\partial\mu_i}{\partial m}=-A_i\mu_iy_2^2(x,\mu_i),
$$
where
$A_i=\left(\int_0^1my_2^2(x,\mu_i)\,dx\right)^{-1}.$

Let us now calculate 
$\frac{\partial\rho_i}{\partial m}.$
Let us multiply~(\ref{dy2}) by $y(x,\mu_i)$ (this is another Floquet
solution for $\mu_i$ defined in the previous section, we will write
simply $y$ instead of $y(x,\mu_i)$),  
subtract $y''=\frac{1}{4}y-\mu my$ multiplied by $\delta y_2$ and
finally integrate. On the L.H.S. we have
$$
\int_0^1(\delta y_2''y-y''\delta y_2)\,dx=%
\int_0^1(\delta y_2'y-y'\delta y_2)'\,dx=(\delta y_2'y-y'\delta y_2)|_0^1.
$$
Remember that $\rho_i=y_2'(1,\mu_i)$ and 
$y(1,\mu_i)=\tilde{\rho}_iy(0,\mu_i)=\tilde{\rho_i}=\frac{1}{\rho_i}.$ We
see that the L.H.S. is equal to 
$\frac{\delta\rho_i}{\rho_i}=\delta\log|\rho_i|.$
On the R.H.S. we obtain
$$
\int_0^1(-\delta\mu my_2y-\mu\delta my_2y)\,dx.
$$
Using the expression for $\delta\mu,$ we obtain
$$
\frac{\partial\log\rho_i}{\partial m}=A_iB_i\mu_iy_2^2(x,\mu_i)-%
\mu_iy_2(x,\mu_i)y(x,\mu_i),
$$
where $B_i=\int_0^1my_2(x,\mu_i)y(x,\mu_i)\,dx.$

Let us now calculate brackets. We will do it using the
lemma from previous section.

Let us prove that $\{\mu_i,\mu_j\}_1=0.$ It is clear if $i=j,$ so let us
suppose that $i\ne j.$ We have
$$
\{\mu_i,\mu_j\}_1=\int_0^1\frac{\partial\mu_i}{\partial m}J%
\frac{\partial\mu_j}{\partial m}\,dx=A_iA_j\mu_i\mu_j%
\int_0^1y_2^2(x,\mu_i)Jy_2^2(x,\mu_j)\,dx.
$$

Let us now remark that if we have two functions $f$ and $g$ such
that $f,$ $f',$ $g$ and $g'$ are equal to zero at points $0$ and $1$
then
\begin{equation}\label{fJg}
\int_0^1fJg\,dx=-\int_0^1gJf\,dx,
\end{equation}
and
\begin{equation}\label{fKg}
\int_0^1fKg\,dx=-\int_0^1gKf\,dx.
\end{equation}
Using these identities and the lemma we have
$$
\mu_i\mu_j\int_0^1y_2^2(\mu_i)Jy_2^2(\mu_j)\,dx=
\mu_i\int_0^1y_2^2(\mu_i)Ky_2^2(\mu_j)\,dx=
$$
$$
=-\mu_i\int_0^1y_2^2(\mu_j)Ky_2^2(\mu_i)\,dx=
-\mu_i^2\int_0^1y_2^2(\mu_j)Jy_2^2(\mu_i)\,dx=
$$
$$
=\mu_i^2\int_0^1y_2^2(\mu_i)Jy_2^2(\mu_j)\,dx.
$$
Since $i\ne j$ and $\mu_i\ne0$ it follows that 
$\int_0^1y_2^2(\mu_i)Jy_2^2(\mu_j)\,dx=0.$ This implies
that $\{\mu_i,\mu_j\}_1=0.$

We will now prove that $\{\mu_i,\log|\rho_j|\}_1=-\mu_i^2\delta_{ij},$
it will imply that $\{\mu_i,f_j\}_1=\delta_{ij}.$
The proof that $\{\mu_i,\log|\rho_j|\}_1=0$ if $i\ne j$ is analogous to the 
proof that $\{\mu_i,\mu_j\}_1=0.$
Let us find $\{\mu_i,\log|\rho_i|\}_1.$ This is equal to
$$
-A_i^2B_i\mu_i^2\int_0^1y_2^2(\mu_i)Jy_2^2(\mu_i)\,dx%
+A_i\mu_i^2\int_0^1y_2^2(\mu_i)Jy_2(\mu_i)y(\mu_i)\,dx.
$$
Using the identity~(\ref{fJg}) we can see that the first term is equal to zero.
Let us drop the index $i$ for simplicity. We have
$$
A\mu^2\int_0^1y_2^2Jy_2y\,dx=A\mu^2\int_0^1y_2^2(mD+Dm)y_2y\,dx=
$$
$$
=A\mu^2\left[\int_0^1y_2^2m(y_2y)'\,dx+\int_0^1y_2^2(my_2y)'\,dx\right]=
$$
$$
=A\mu^2\left[\int_0^1y_2m(y_2y)'\,dx-\int_0^1my_2y(y_2^2)'\,dx\right]=
$$
$$
=A\mu^2\int_0^1my_2^2(y_2y'-y_2'y)\,dx.
$$
The expression
$y'y_2-yy_2'$ is a Wronskian $W(y,y_2).$ It is a constant, not depending on 
$x.$ Let us calculate it for $x=1.$ Since 
$y(1,\mu_i)=\tilde{\rho}_iy(0)=\frac{1}{\rho_i}=\frac{1}{y_2'(1,\mu_i)},$
we obtain $y_2y'-y_2'y=-1.$ This implies (remember the definition of $A$)
$$
A\mu^2\int_0^1my_2^2(y_2y'-y_2'y)\,dx=-A\mu^2\int_0^1my_2^2\,dx=-\mu^2.
$$
Hence, we obtain $\{\mu_i,\log|\rho_j|\}_1=-\mu_i^2\delta_{ij}.$

If we prove that $\{\log|\rho_i|,\log|\rho_j|\}_1=0,$ it will imply that 
$\{f_i,f_j\}_1=0.$ But the proof that $\{\log|\rho_i|,\log|\rho_j|\}_1=0$ is
analogous to previous calculations.

This finishes the proof. $\Box$

\noindent{\bf Theorem 2.} The variables $\mu_i$ and $g_j$ 
are conjugate with respect to the second bracket:
$$
\{\mu_i,\mu_j\}_2=0,\quad
\{\mu_i,g_j\}_2=\delta_{ij},\quad
\{g_i,g_j\}_2=0.
$$

\noindent{\bf Proof} is analogous to the proof of Theorem~1. $\Box$

\section*{Acknowledgments}

The author is very grateful to the Centre de Recherches Math\'ematiques (CRM)
for its hospitality. The author would like to thank Prof. A.~Broer 
and Prof. P.~Winternitz for useful discussions.


\begin{thebibliography}{99}
\bibitem{FM} H.~Flaschka, D.~W.~McLaughlin, {\em Canonically conjugate 
variables for the Korteweg-de Vries equation and the Toda lattice with 
periodic boundary conditions.} Progr. Theoret. Phys. 55 (1976), no.~2, 
438--456.
\bibitem{VN1} A.~P.~Veselov, S.~P.~Novikov, {\em On Poisson brackets 
compatible with algebraic geometry and Korteweg-de~Vries dynamics
on the set of finite-zone potentials.} (Russian) Dokl. Akad. Nauk SSSR 266 
(1982), no.~3, 533--537. Translation in Soviet Math. Dokl. 26 (1982),
no.~2, 357--362.
\bibitem{VN2} A.~P.~Veselov, S.~P.~Novikov, {\em Poisson brackets and complex 
tori.} (Russian) Algebraic geometry and its applications. Trudy Mat. Inst. 
Steklov. 165 (1984), 49--61.
\bibitem{KP1} I.~M.~Krichever, D.~H.~Phong, {\em On the integrable geometry 
of soliton equations and $N=2$ supersymmetric gauge theories.} J.~Differential
Geom. 45 (1997), no. 2, 349--389.
\bibitem{KP2}I.~M.~Krichever, D.~H.~Phong, {\em Symplectic forms in the 
theory of solitons.} In {\em Surveys in differential geometry: integral 
systems [integrable systems],} 239--313, Surv. Differ. Geom., IV, Int. Press, 
Boston, MA, 1998. 
\bibitem{CH} R. Camassa, D. Holm, {\em An integrable shallow water equation
with peaked solutions.} Phys. Rev. Lett. 71 (1993), no.~11, 1661-1664.
\bibitem{CM} A.~Constantin, H.~P.~McKean, {\em A shallow water equation
on the circle.} Comm. Pure Appl. Math. 52 (1999), no.~8, 949--982.
\bibitem{CE} A.~Constantin, J.~Escher, {\em Well-posedness, global
existence, and blowup phenomena for a periodic quasi-linear
hyperbolic equation.} Comm. Pure Appl. Math. 51 (1998), no.~5, 475--504.
\bibitem{CMo} A.~Constantin, L.~Molinet, {\em Global weak solutions
for a shallow water equation.} Comm. Math. Phys. 211 (2000), no.~1, 45--61.
\bibitem{ACFHM} M.~S.~Alber, R.~Camassa, Yu.~N.~Fedorov, D.~D.~Holm,
J.~E.~Marsden, {\em The complex geometry of weak piecewise smooth solutions 
of integrable nonlinear PDE's of shallow water and Dym type.} 
Comm. Math. Phys. 221 (2001), no.~1, 197--227.
\bibitem{GH} F.~Gesztesy, H.~Holden, {\em Algebro-geometric solutions of the 
Camassa-Holm hierarchy.} Rev. Mat. Iberoamericana 19 (2003), no.~1, 73--142.
\end{thebibliography}
\end{document}